\documentstyle[epsfig,twocolumn,seceq]{jpsj}

\def\runtitle{Oblique triangular spin structure in CsCu$_{1-x}$Co$_x$Cl$_3$}
\def\runauthor{Ono {\it et al.}}

\title
{
Oblique triangular antiferromagnetic phase \\
in CsCu$_{1-x}$Co$_x$Cl$_3$
}

\author
{ 
T. Ono, T. Kato\footnote{Present address: Faculty of Education, Chiba University, 263-8522 Chiba, Japan.}, H. Tanaka, A. Hoser$^{\dag}$, N. St\"{u}\ss er$^{\dag}$ and U. Schotte$^{\dag}$
}

\inst
{
Department of Physics, Tokyo Institute of Technology, 152-8551 Tokyo, Japan\\
$^{\dag}$Hahn-Meitner-Institut, Glienicker Stra\ss e 100, D-14109 Berlin, Germany
}

\recdate
{
\today
}

\abst
{
	The spin-$\frac{1}{2}$ stacked triangular antiferromagnet CsCu$_{1-x}$Co$_x$Cl$_3$ with $0.015<x<0.032$ undergoes two phase transitions at zero field. The low-temperature phase is produced by the small amount of Co$^{2+}$ doping. In order to investigate the magnetic structures of the two ordered phases, the neutron elastic scattering experiments have been carried out for the sample with $x\approx 0.03$. It is found that the intermediate phase is identical to the ordered phase of CsCuCl$_3$, and that the low-temperature phase is an oblique triangular antiferromagnetic phase in which the spins form a triangular structure in a plane tilted from the basal plane. The tilting angle which is 42$^{\circ}$ at $T=1.6$\,K decreases with increasing temperature, and becomes zero at $T_{\rm N2} =7.2$\,K. An off-diagonal exchange term is proposed as the origin of the oblique phase.
}

\begin{document}
%{\sf 75.25.+z, 75.30.Kz}
\sloppy
\maketitle

\section{Introduction}
	Among the group of antiferromagnets which have a hexagonal ABX$_3$ triangular structure, many experimental and theoretical investigations for CsCuCl$_3$ have been done by several authors\,\cite{Nojiri98,Werner97,Schotte98,Jacobs98}, due to the remarkable features of this compound. 
Since the Cu$^{2+}$ ion is Jahn-Teller active, CsCuCl$_3$ undergoes a structural phase transition at $T_{\rm t}=423$\,K, where Jahn-Teller distorted octahedra CuCl$_6$ order so that their longest axes form a helix along the  $c$-axis with a repeatlength of six\,\cite{Schlueter66,Kroese74,Hirotsu77}. The low-symmetric crystal structure produces the antisymmetric interaction of the Dzyaloshinsky-Moriya (D-M) type between the adjacent spins in the chain with the $D$-vector parallel to the $c$-direction. The exchange interactions along and between the chains are ferromagnetic and antiferromagnetic, respectively, and their values have been evaluated as $J_0/k_{\rm B}=28$\,K and $J_1/k_{\rm B}=-4.9$\,K\,\cite{Tazuke81,Tanaka92} in the definition of ${\cal H}=-\sum_{<i,j>}2J_{ij}S_i\cdot S_j$. The magnetic phase transition occurs at $T_{\rm N}=10.5$\,K\,\cite{Adachi80,Weber96}. In the ordered state, spins lie in the basal plane and form the 120$^{\circ}$-structure, while along the $c$-direction, a long period (about 71 triangular layers) helical incommensurate arrangement is realized, due to the competition between the ferromagnetic interaction and the D-M interaction along the chain. \par
\begin{figure}[h]
  \begin{center}
     \epsfxsize=8cm
	\epsfbox{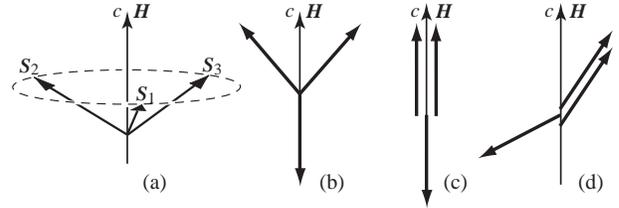}
\end{center}
\caption{Spin structures of 2D TAF in the magnetic field. (a) Umbrella-type structure, (b) low-field coplanar structure, (c) collinear structure and (d) high-field coplanar structure.}
\label{umbrella}
\end{figure}
\begin{figure}[htbp]
  \begin{center}
     \epsfxsize=6.5cm
	\epsfbox{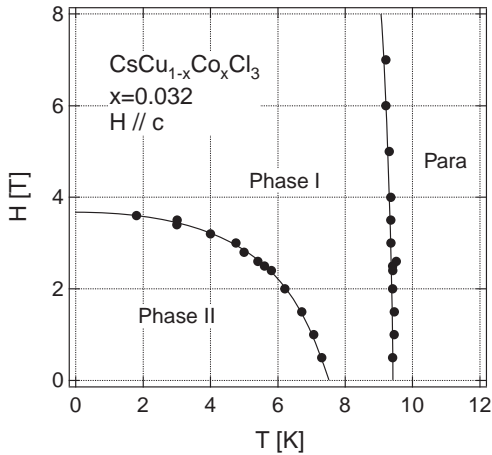}
\end{center}
\caption{The magnetic phase diagram of CsCu$_{1-x}$Co$_x$Cl$_3$ with $x=0.32$ for the magnetic field parallel to the $c$-axis.}
\label{Ex_diagram}
\end{figure}

Since the magnitude of spin on Cu$^{2+}$ is $S=\frac{1}{2}$, quantum fluctuations are important when one describes the magnetic behavior in the magnetic field\,\cite{Nikuni93}. Particularly, the magnetization process with a strong field often reveals quantum effects as macroscopic phenomena. Using quantum Monte Carlo calculations or spin wave theory, for the 2D triangular antiferromagnet (TAF) in the isotropic limit, it is predicted that the magnetization plateau appears at around one third of the saturation value $M_{\rm s}$\,\cite{Nishi86,Chub91}.
 According to these predictions, when the external field ${\mib H}$ is applied parallel to the $c$-axis, 2D TAF undergoes a successive phase transition as shown in Fig.\,\ref{umbrella}\,(b) low-field coplanar, (c) collinear and (d) high-field coplanar structure in that order. In the field region where the magnetization plateau appears, the spin structure corresponds to (c) structure. %
\vspace{5mm}
For CsCuCl$_3$, instead of a plateau, there is a small magnetization jump around $M_{\rm s}/3$, when the field is applied parallel to the $c$-axis.
 This fact is understood as the competition between weak planar anisotropy and quantum fluctuation, i.e., Fig.\,\ref{umbrella}\,(a) structure is stabilized by the planar anisotropy up to $H_{\rm c}=$ 12.5\,T $>H_{\rm s}/3$ and Fig.\,\ref{umbrella}\,(d) structure is stabilized by quantum fluctuation above $H>H_{\rm c}$ with skipping (b) and (c) structures.
  The existence of the plateau at $M=M_{\rm s}/3$ is strongly depend on the magnetic anisotropies. Recently, Honecker {\it et\,al.}\,\cite{Honecker99} argued that the proper combination for the interchain interaction $|J_1|/J_0$ and $XXZ$ anisotropy $\Delta=J^{z}/J^{xy}$ allows the existence of the plateau at $M=M_{\rm s}/3$, even in the region of $\Delta <1$ (small planar anisotropy). Here $J^{\alpha}(\alpha=z\ {\rm and}\ xy)$ are anisotropic exchange constants defined by ${\cal H}=-2\sum_i [J_{i\,i+1}^{zz}S_i^z S_{i+1}^z+\frac{1}{2}J_{i\,i+1}^{xy}(S_i^+S_{i+1}^- + S_i^-S_{i+1}^+)]$. Therefore, it is very interesting to study how the phase transition changes, when the macroscopic anisotropy of CsCuCl$_3$ is controlled. Because of this motivation, we have synthesized the mixture system CsCu$_{1-x}$Co$_x$Cl$_3$ with few percents of Co$^{2+}$ ion in the concentration $x$, since CsCoCl$_3$ is known as a pseudospin-$\frac{1}{2}$ TAF with easy axis anisotropy\,\cite{Mekata78}. In our previous paper \cite{Ono2000}, we reported the results of the magnetic measurements for CsCu$_{1-x}$Co$_x$Cl$_3$ with $x<0.032$ and the magnetic phase diagram for temperature versus magnetic field up to 7\,T. Figure \ref{Ex_diagram} shows the magnetic phase diagram for the external field versus temperature for the sample with $x=0.032$ in the field parallel to the $c$-axis.
\vspace{5mm}
Before the magnetization measurements, our first speculation was that when Co$^{2+}$ ions are doped to reduce the planar anisotropy Fig.\,\ref{umbrella}\,(a) structure is realized in the ground state at zero field, and (b) structure which benefits quantum fluctuation might appear in the field region $H\leq H_{\rm s}/3$.\par
On the other hand, the experimental results of the magnetization measurements indicate that the spin structure of phase II is stabilized with the help of the axial anisotropy. With decreasing Co$^{2+}$ concentration $x$, the area of phase II is reduced and then disappears when $x\leq 0.005$. Thus it is reasonable to infer that, the spin structure in the intermediate phase I is identical to that of CsCuCl$_3$. However, within the magnetic measurements the spin structure of the intermediate phase and low-temperature phase cannot be determined. Therefore, in order to determine the spin structure of these two phases, we have performed neutron scattering experiments for the sample with $x\approx 0.03$.

\section{Equipment}
	Single crystals of CsCu$_{1-x}$Co$_x$Cl$_3$ were prepared by the Bridgman method from the melt of the single crystals of CsCuCl$_3$ and CsCoCl$_3$. The details of the sample preparation are described in the previous paper \cite{Ono2000}.\par
Neutron scattering experiments were performed at the BER II Research Reactor of the Hahn-Meitner-Institute using the triple axis spectrometer E1 operated in the double-axis mode. The incident neutron energy was fixed at $E_i=13.9$\,meV by Bragg reflection at the pyrolytic graphite monochromator, and the horizontal collimations are chosen as 40'-40'-40'-open. A single crystal of CsCu$_{1-x}$Co$_x$Cl$_3$ with $x\approx 0.03$ was used. We label this sample CsCu$_{0.97}$Co$_{0.03}$Cl$_3$. The sample has the shape of triangular prism 5\,mm in width and 15\,mm in length. The sample was mounted in the cryostat with its [1, 1, 0] and [0, 0, 1] axes in the scattering plane.

\section{Results}

From the previous magnetic measurements, for the sample with CsCu$_{0.97}$Co$_{0.03}$Cl$_3$, undergoes phase transitions at $T_{\rm N1}\simeq 9.5$\,K and $T_{\rm N2}\simeq 7.2$\,K. First, we confirmed that the crystal structure of the present system is identical to that of pure CsCuCl$_3$, i.e., the nuclear Bragg peaks were observed at $(00\ell)$ with $\ell=0\sim 6$, which are indicative of the six-fold helix. From the nuclear Bragg peaks at (1, 1, 0) and (0, 0, 6), the lattice constants at helium temperature were determined as $a=7.137$\,\AA\ and $c=18.036$\,\AA. \par
\begin{figure}[h]
  \begin{center}
     \epsfxsize=7cm
	\epsfbox{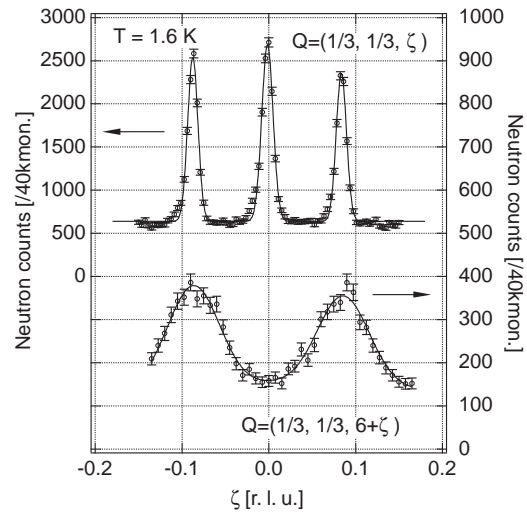}
\end{center}
\caption{Scans along $Q=(1/3, 1/3, \zeta)$ with (a) $\zeta=0$ and (b) $\zeta=6$. Solid lines are the results of Gaussian fitting.}
\label{profiles_16k}
\end{figure}

Figure \ref{profiles_16k} shows the scans along $Q=(1/3, 1/3, \zeta)$ and $(1/3, 1/3, 6+\zeta)$ at $T=1.6$\,K in the phase II. The magnetic Bragg scattering is composed of one central peak at $\zeta=0$ and two side peaks at $\zeta=\pm\delta$ with $\delta =0.084$.%
\vspace{5mm}

\begin{figure}[h]
  \begin{center}
     \epsfxsize=7cm
	\epsfbox{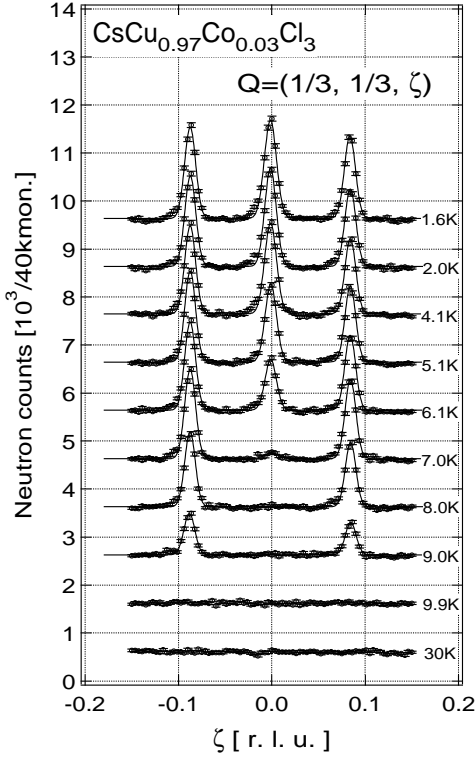}
\end{center}
\caption{Scans along the $c^*$-axis around $Q=(1/3, 1/3, 0)$ at various temperatures.}
\label{T_profiles}
\end{figure}

\begin{figure}[htb]
	\begin{center}
	\epsfxsize=7cm
	\epsfbox{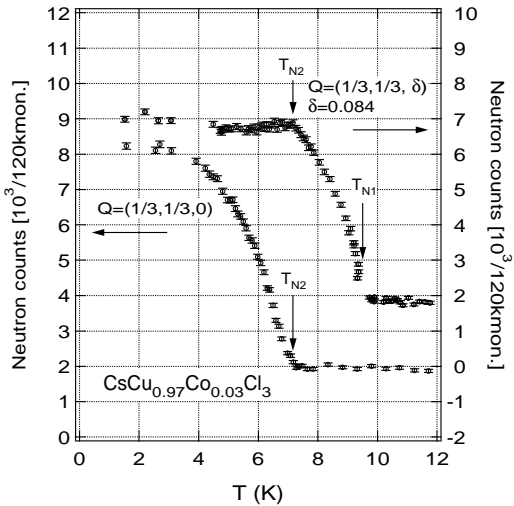}
\end{center}
\caption{Magnetic Bragg intensities at $Q=(1/3, 1/3, 0)$ and $(1/3, 1/3, 0.084)$ as a function of temperature. N\'{e}el temperatures are indicated by arrows.}
\label{PeakVar}
\end{figure}

 The temperature dependence of the scan profile (see Fig.\,\ref{T_profiles}) shows that possible changes of $\delta$ is negligible, similar to pure CsCuCl$_3$.%
The possibility that the central peak and the side peaks come from different magnetic domains is excluded, because the temperature variations of their intensities are strongly correlated as shown in Fig.\,\ref{PeakVar}. %
%The difference between the intensities of two side peaks is due to small misalignment of the sample.%
 We see that the magnetic unit cell is enlarged by $\sqrt{3}\times\sqrt{3}$ in the basal plane, which is characteristic of the triangular spin arrangement. The side peaks indicate that the helical spin  structure is realized along the $c$-axis. The value of $\delta$ corresponds to the period of the helix. The period is evaluated as $6/\delta =71.4$ Cu-sites, which is almost the same as that in CsCuCl$_3$, {\it i.e.}, 70.6 Cu-sites\,\cite{Adachi80}.\par
\begin{figure}[h]
  	\begin{center}
     \epsfxsize=7cm
	\epsfbox{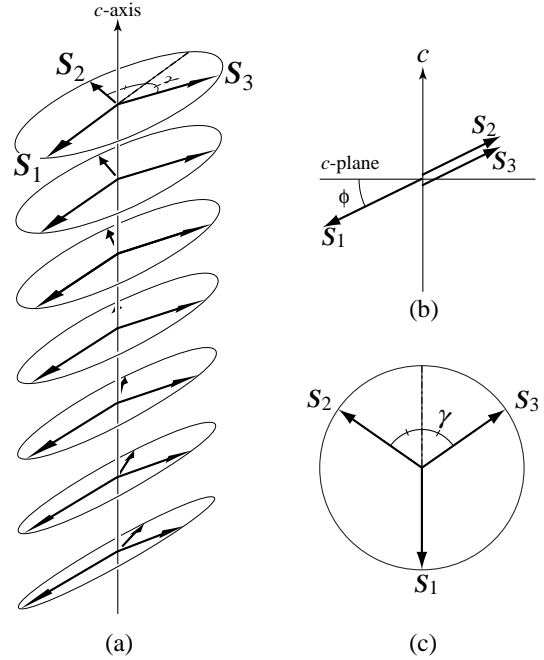}
\end{center}
\caption{(a)\,Oblique triangular antiferromagnetic structure in the low-temperature phase. Tilted spin planes represented by circles are stacked along the $c$-axis with forming a helix with a pitch 5 degrees. (b)\,The angle $\phi$ indicates the tilt between the spin plane and the $c$-plane. (c)\,The angle $\gamma$ is half of the angle between $S_2$ and $S_3$.}
\label{disks}
\end{figure}

It is remarkable that in the present system, the magnetic Bragg peak is observed at $Q=(1/3, 1/3, 0)$, which is absent in CsCuCl$_3$. On the other hand, around (1/3, 1/3, 6), the central peak is very weak as compared with the side peaks (see Fig.\,\ref{profiles_16k}). Since the angle between the scattering vector and the $c^*$-axis is 90$^{\circ}$ for $Q=(1/3, 1/3, 0)$ and 15.7$^{\circ}$ for $Q=(1/3,1/3, 6)$, the $c$-axis component of spins is much less visible around the (1/3, 1/3, 6) reflection than around the (1/3, 1/3, 0) reflection. On the other hand, the basal plane component of spins make a larger contribution to the magnetic reflections around (1/3, 1/3, 6). Thus, it is deduced that the central peaks originate from the $c$-axis component of spins, while the basal plane component gives rise to  the side peaks as for CsCuCl$_3$. We refer to a plane spanned by the spins on a triangular lattice as {\it spin-plane} in this paper. The present result leads us straightforward to conclude that the phase II is an oblique triangular phase, in which the spin-plane is tilted with an angle $\phi$ from the basal plane (see Fig.\,\ref{disks}\,(a) and (b)).%
\vspace{5mm}
 The spin-planes are stacked along the $c$-axis with forming a helix with a pitch $360\times \delta /6 \sim 5$\,degrees. \par
\begin{figure}[h]
  \begin{center}
     \epsfxsize=7cm
	\epsfbox{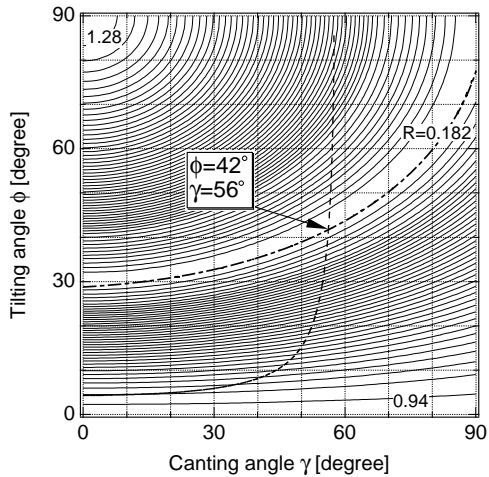}
\end{center}
\caption{Contour plot of the reliability factor $R$ as a function of the angle $\gamma$ and $\phi$. Dot dashed line indicates the minimized $R$-factor $R=0.182$. Dashed line indicates the line which satisfies eq.\,(\ref{spon}).}
\label{contour}
\end{figure}

It is quite natural to consider that the spin structure in each spin-plane is close to the 120$^{\circ}$-structure, because there is no anisotropy which is strong enough to deform largely the 120$^{\circ}$-structure or lift the moments from the spin-plane. Although the D-M interaction is of the same order in magnitude as $J_1$, its effect is reduced to be $\sim D^2/J_0$ due to the competition with the intrachain exchange interaction $J_0$\,\cite{Tanaka92}. In order to take account of a slight deformation from the 120$^{\circ}$-structure, the angle $2\gamma$ between the spins ${\mib S}_2$ and ${\mib S}_3$, is adopted as a variable parameter as illustrated in Fig.\,\ref{disks}\,(c). Under the assumption that the thermal averages of spins on all sublattices are the same, the integrated intensity of the magnetic Bragg peaks for $Q=(h, h, \zeta)$ with $h=n/3$ is given by%
\begin{eqnarray}
I &\propto&|f(Q)|^2L(\theta)|F(h, h, \zeta)|^2\\
L(\theta)&=& \cos \left(\frac{\pi}{2}-\theta-\arccos\sqrt{\frac{a^2\zeta ^2}{4c^2h^2+a^2\zeta^2}}\right)\nonumber
\label{eq_intensity}
\end{eqnarray}%
and $I=0$ for integer $h$, where $\theta$ is half of the scattering angle, $L(\theta)$ is the Lorentz factor for the parallel scans, $f(Q)$ is the magnetic form factor and $a$ and $c$ are lattice constants. The values of $f(Q)$ were taken from reference\,\cite{Watson83}. $F(h, h, \zeta)$ is the magnetic structure factor. Due to the long-period helical spin structure along the $c$-axis, the component in the $a$-$b$-plane and the $c$-axis component of the spins give Bragg peaks at $\zeta =\pm\delta$ and $\zeta =0$, respectively. The structure factors for the planar component $F_{ab}$ and the axial component $F_c$ are described as%
\begin{eqnarray}
|F_{ab}(hh\zeta)|^2	&=&d_{(hh\zeta)}^2\left\{2\left(\frac{h}{a}\right)^2+\left(\frac{\zeta}{c}\right)^2\right\}\nonumber\\
&\times &\{(1+\cos\gamma)^2\cos^2\phi+3\sin^2\gamma\}\\
|F_c(hh\zeta)|^2	&=&d_{(hh\zeta)}^2\left(\frac{2h}{a}\right)^2(1+\cos^2\gamma)\sin^2\phi,
\label{st_factors}
\end{eqnarray}
where $d_{(hh\zeta)}$ is the spacing of the $(h, h, \zeta)$ lattice plane. Integrated Bragg intensities are calculated for various angles of $\phi$ and $\gamma$. In Fig.\,\ref{contour} the variation of the reliability factor $R$ given by
\[
R=\sum_{h, h, \zeta}\bigl|I_{\rm calc}(h, h, \zeta)-I_{\rm obs}(h, h, \zeta)\bigr|\Bigg/ \sum_{h, h, \zeta}I_{\rm obs}(h, h, \zeta).
\]
is represented as a contour map.%
\vspace{5mm}
 We see that the $R$-factor has the same minimum value on the dot-dashed line in Fig.\,\ref{contour}. With decreasing tilting angle $\phi$, the canting angle $\gamma$ which gives minimum $R$-factor rapidly decreases, and become zero when $\phi=29.2^{\circ}$. Since the magnetic structure could not be determined uniquely only from the magnetic Bragg intensities, the value of the spontaneous magnetization along the $c$-axis detected from magnetization measurements (see Fig.\,2 on ref.\,\cite{Ono2000}) is necessary to give the conclusive answer. For the oblique triangular structure, each sublattice spin is described as
\begin{eqnarray}
	S_1&=&S(\cos\phi , 0, -\sin\phi)		\nonumber\\
	S_2&=&S(-\cos\gamma\cos\phi , -\sin\gamma , \cos\gamma\sin\phi)	\nonumber\\
	S_3&=&S(-\cos\gamma\cos\phi , \sin\gamma , \cos\gamma\sin\phi ).
\end{eqnarray}
Hence, the spontaneous magnetization $M_{\rm sp}$ per site is given by
\begin{equation}
	M_{\rm sp}=\frac{2\cos\gamma-1}{3}\sin\phi\ [\mu_{\rm B}/\textrm{magnetic ion}].
	\label{spon}
\end{equation}
In Fig.\,\ref{contour}, the dashed line gives $M_{\rm sp}=0.025\,\mu _{\rm B}$ which was observed by the previous magnetization measurements\,\cite{Ono2000}. From the intersection of dashed and dot-dashed lines, the angles $\phi$ and $\gamma$ are obtained as $(\phi, \gamma)=(42^{\circ}, 56^{\circ})$. \par
In Table \ref{Bragg}, we show the experimental and calculated intensities of the Bragg peaks around $Q=(1/3, 1/3, 0)$, (1/3, 1/3, 6) and (2/3, 2/3, 0) together with the calculated intensities which corresponds to the \textsf{A}- and \textsf{B}-point in Fig.\,\ref{contour}. %
\begin{fulltable}[bt]
\caption{Experimental and calculated intensities of several magnetic Bragg reflections in CsCu$_{1-x}$Co$_x$Cl$_3$ with $x=0.03$ at $T=1.6$\,K. They are normalized to the total peak intensities around $Q=(1/3, 1/3, 0)$ reflection. The calculated intensities at \textsf{A}- and \textsf{B}-point in Fig.\,\ref{contour} are also shown for comparison.}
	\begin{fulltabular}{@{\hspace{\tabcolsep}\extracolsep{\fill}}ccccc}
	%\begin{fulltabular}{ccccc} % In second brace, l = left, r = right,
% c = centered and d = decimal justification.

\hline
$(h, h, \zeta)$ &$I_{\rm obs}$ &$I_{\rm cal}\pmatrix{\phi =42^{\circ}\\ \gamma =57^{\circ}}$ &$I_{\rm cal}(\textsf{A}) \pmatrix{\phi =0^{\circ}\\ \gamma =60^{\circ}}$ &$I_{\rm cal}(\textsf{B}) \pmatrix{\phi =90^{\circ}\\ \gamma =57.49^{\circ}}$\\
\hline
%\tableline % Creates a horizontal line.
$\left(\frac{1}{3}, \frac{1}{3}, -\delta\right)$& 0.324& 0.306& 0.497& 0.155\\
$\left(\frac{1}{3}, \frac{1}{3}, 0\right)$& 0.384& 0.384& 0& 0.689\\
$\left(\frac{1}{3}, \frac{1}{3}, \delta\right)$& 0.292& 0.310& 0.503& 0.156\\
[2mm]
$\left(\frac{1}{3}, \frac{1}{3}, 6-\delta\right)$& 0.210& 0.230& 0.373& 0.116\\
$\left(\frac{1}{3}, \frac{1}{3}, 6\right)$& 0.012& 0.011& 0& 0.019\\
$\left(\frac{1}{3}, \frac{1}{3}, 6+\delta\right)$& 0.186& 0.226& 0.366& 0.114\\
[2mm]
$\left(\frac{2}{3}, \frac{2}{3}, -\delta\right)$& 0.171& 0.266& 0.432& 0.135\\
$\left(\frac{2}{3}, \frac{2}{3}, 0\right)$& 0.252& 0.330& 0& 0.573\\
$\left(\frac{2}{3}, \frac{2}{3}, +\delta\right)$& 0.172& 0.266& 0.432& 0.134\\
[2mm] & & $R=0.182$ & $R=0.947$ & $R=0.598$\\
[2mm]
$M_{\rm sp}$\,[$\mu_{\rm B}$/ion]&0.025 &0.025& 0& 0.025\\
\hline
\end{fulltabular}
\label{Bragg}
\end{fulltable}%
Experimental and calculated intensities are normalized for the total intensities of central and side peaks around (1/3, 1/3, 0). The spin structure of CsCuCl$_3$ correspond to the case $\phi=0^{\circ}$ and $\gamma=60^{\circ}$ which is indicated by \textsf{A} in Fig.\,\ref{contour}. We see from Table \ref{Bragg} that the spins form the triangular structure neither in the $c$-plane, nor in the plane including the $c$-axis, but in a plane tilted from the $c$-plane with the angle $\phi=42^{\circ}$ and that the triangular spin structure ($\gamma=56^{\circ}$), is close to the 120$^{\circ}$ spin structure. The difference between the observed and calculated values of the scattering intensities may come from the fact that we did not taking into account of the variation of the magnitude of $S$ for each sublattice.\par
As shown in Fig.\,\ref{PeakVar}, with increasing temperature, the intensity of the central peak decreases and disappears at $T_{\rm N2}=7.2$\,K. On the other hand, the intensity of the side peak is almost constant up to $T_{\rm N2}$, and decreases rapidly, and becomes zero at $T_{\rm N1}=9.5$\,K. Both phase transitions look sharp, which indicates a good homogeneity of the sample. From the present result, we see that the tilting angle $\phi$ decreases with increasing temperature, and becomes zero at $T_{\rm N2}$. Thus, we can deduce that the phase I between $T_{\rm N2}$ and $T_{\rm N1}$ is identical to the ordered phase of CsCuCl$_3$, which is consistent with the fact that the Co$^{2+}$ doping produces the phase II in the low-temperature and low-field region in the ordered state of CsCuCl$_3$, and its area increases with increasing the Co$^{2+}$ concentration $x$.\par
CsCuCl$_3$ has a weak planar anisotropy, while CsCoCl$_3$ is an Ising spin system\,\cite{Collins97}. Phase transitions in the random spin systems with competing anisotropies have been investigated in Fe$_{1-x}$Co$_x$Cl$_2$\,\cite{Wong83} and Fe$_{1-x}$Co$_x$Cl$_2\cdot$2H$_2$O\,\cite{Katsu85} by means of neutron scattering. Both systems  undergo two phase transitions, which are characterized by the orderings of two spin components. In Fe$_{1-x}$Co$_x$Cl$_2$, the ordering of one spin component is drastically altered by the ordering of the other component, so that the lower-temperature transition is rather smeared. On the other hand, in Fe$_{1-x}$Co$_x$Cl$_2\cdot$2H$_2$O, two phase transitions  are sharp, and two order parameters are decoupled. CsCu$_{0.97}$Co$_{0.03}$Cl$_3$ differs from the both in the ordering nature. In the present system, the basal plane component of spins orders first at $T_{\rm N1}$ and then the $c$-axis component orders at $T_{\rm N2}$, so that the oblique triangular spin structure is realized. Both phase transitions are fairly sharp, and both components are not decoupled, {\it i.e.}, the growth of the basal plane component of spins below $T_{\rm N2}$ is suppressed due to the growth of the $c$-axis component which is equivalent to the increase of tilting angle $\phi$. The present system is the first example that undergoes the definite phase transition to the oblique spin phase. Its mechanism seems not to be explained by only the competing anisotropies, because mixed triangular antiferromagnetic systems with axial and planar anisotropies such as Rb$_{1-x}$K$_x$NiCl$_3$\,\cite{Tanaka93} and CsMn(Br$_x$I$_{1-x}$)$_3$\,\cite{Ono98,Ono99} exhibited neither mixed ordered phase nor oblique phase. \par
In CsCuCl$_3$, the off-diagonal exchange term of the form
\begin{equation}
{\cal H}_{od}=J_{ij}^{zx}(S_i^zS_j^x+S_i^xS_j^z)+J_{ij}^{zy}(S_i^zS_j^y+S_i^yS_j^z)
\end{equation}
is allowed, because the elongated axes of CuCl$_6$ octahedra are canted from the $c$-axis, so that the local symmetry is lower than the trigonal one. Here we take the $z$-axis parallel to the $c$-axis and $x$- and $y$-axes in the basal plane. The off-diagonal exchange term has been proposed to interpret the nature of the mixed ordered state in Fe$_{1-x}$Co$_x$Cl$_2$\,\cite{Wong83,Muka81}. Recently, Pleimling\,\cite{Pleim99} discussed the ordering of the perpendicular spin component in the Ising-like system FeBr$_2$ on the basis of the off-diagonal exchange term.  Due to the off-diagonal term, the ordering of the basal plane component of spins can induce the ordering of the $c$-axis component. We suggest that the doping of small amounts of Co$^{2+}$ ions suppresses the planar anisotropy in the average, so that the effect of the off-diagonal term is relatively enhanced. However, whether the orderings of both components occur simultaneously or separately as observed in the present system is unclear. This problem should be investigated theoretically.

\section{Conclusion}

	We have performed elastic neutron scattering experiments on the triangular antiferromagnetic system CsCu$_{0.97}$Co$_{0.03}$Cl$_3$ which undergoes two phase transitions at $T_{\rm N1}=9.5$\,K and $T_{\rm N2}=7.2$\,K. It was found that the low-temperature phase is an oblique triangular antiferromagnetic phase with the spin-plane tilted from the basal plane, and that the intermediate phase is identical to the ordered phase of CsCuCl$_3$. We suggest that the off-diagonal exchange interaction gives rise to the oblique phase in the present system.

\section*{Acknowledgments}

The authors wish to thank K. D. Schotte and H. Shiba for useful dicussions.

\end{document}